\documentclass[comsoc,conference,draftclsnofoot, onecolumn]{IEEEtran}
\usepackage{cite}
\usepackage[pdftex]{graphicx}
\usepackage{amsmath}
\usepackage{subcaption}
\usepackage{bbm}
\usepackage{xcolor}
\usepackage[ruled, linesnumbered, commentsnumbered, longend]{algorithm2e}
\let\oldnl\nl% Store \nl in \oldnl
\newcommand{\nonl}{\renewcommand{\nl}{\let\nl\oldnl}}% Remove line number for one line

\newcommand{\subjto}{\textup{subject to}}

\newcounter{problem}
\newcounter{save@equation}
\newcounter{save@problem}

\makeatletter
\newenvironment{problem}
 {\setcounter{problem}{\value{save@problem}}%
  \setcounter{save@equation}{\value{equation}}%
  \let\c@equation\c@problem
  \subequations
  }
 {\endsubequations
  \setcounter{save@problem}{\value{equation}}%
  \setcounter{equation}{\value{save@equation}}%
 }

\DeclareMathOperator*{\Minimize}{minimize}

\begin{document}
\bstctlcite{IEEEexample:BSTcontrol}
\title{A Deep Reinforcement Learning Approach for Service Migration in MEC-enabled Vehicular Networks}

\author{
\IEEEauthorblockN{Amine Abouaomar\IEEEauthorrefmark{1}\IEEEauthorrefmark{3}, Zoubeir Mlika\IEEEauthorrefmark{1}, Abderrahime Filali\IEEEauthorrefmark{1}, Soumaya Cherkaoui\IEEEauthorrefmark{1}, and Abdellatif Kobbane\IEEEauthorrefmark{3}}
\IEEEauthorblockA{\IEEEauthorrefmark{1}INTERLAB, Engineering Faculty, Université de Sherbrooke, Sherbrooke (QC), Canada}
\IEEEauthorblockA{\IEEEauthorrefmark{2}ENSIAS, Mohammed V University, Rabat, Morocco}

% Emails : amine.abouaomar@usherbrooke.ca, zoubeir.mlika@usherbrooke.ca , abderrahime.filali@usherbrooke.ca \\ abdelltif.kobbane@um5.ac.ma, soumaya.cherkaoui@usherbrooke.ca
}

\maketitle

\begin{abstract}
Multi-access edge computing (MEC) is a key enabler to reduce the latency of vehicular network. Due to the vehicles mobility, their requested services (e.g., infotainment services) should frequently be migrated across different MEC servers to guarantee their stringent quality of service requirements. In this paper, we study the problem of service migration in a MEC-enabled vehicular network in order to minimize the total service latency and migration cost. This problem is formulated as a nonlinear integer program and is linearized to help obtaining the optimal solution using off-the-shelf solvers. Then, to obtain an efficient solution, it is modeled as a multi-agent Markov decision process and solved by leveraging deep Q learning (DQL) algorithm. The proposed DQL scheme performs a proactive services migration while ensuring their continuity under high mobility constraints. Finally, simulations results show that the proposed DQL scheme achieves close-to-optimal performance.
\end{abstract}


\IEEEpeerreviewmaketitle

\section{Introduction}

Intelligent transportation systems (ITS) represents a critical component of the Internet of Things (IoT) and future smart cities \cite{7835337}. ITS will potentially provide a more secure transportation environment through effective vehicle coordination and efficient resource management \cite{8647858, 9497103, triwinarko2021phy, 7585028}. In addition to safety, the ITS ecosystem will provide entertainment services such as video streaming and gaming, which can be extended to in-vehicle augmented reality \cite{7840359, 9003407, 8004158}. To achieve these promising features, vehicles must be able to communicate, exchange information and access given services with low latency. Therefore, vehicles must operate in an environment that meets these requirements.

Multi-access edge computing (MEC) is envisioned as a key component for fifth-generation (5G) ultra-reliable low-latency communications (uRLLC) services, alongside software-defined networking (SDN)~\cite{9326402} technology. On the one hand, MEC can be leveraged as an emerging computational paradigm that provides efficient computational capabilities to vehicles deployed in close proximity to MEC servers while ensuring a low latency. On the other hand, SDN technology enables seamless, transparent, and efficient control through the separation of the data plane and the control plane, which simplifies network operation and management \cite{filali2020preemptive, 8450284}. Therefore, a MEC-enabled vehicular network can benefit from SDN to provide efficient resource management and uRLLC vehicular services \cite{9318243, abouaomar2021service, abouaomar21globecom, 8290681}. Nevertheless, due to the limited resources of MECs and the high mobility of vehicles, there are many challenges. In particular, the requested vehicular services must be located and migrated to different MEC servers to guarantee their continuity \cite{8737560,8629587,8463562}.

To address these challenges, we investigate the service placement and migration problem in a MEC-enabled vehicular network. We leverage SDN technologies to have efficient control on the MEC servers operations, with the objective of reducing the average service latency of the vehicles. We first, formulate the problem of service placement and migration as a nonlinear integer program that we linearize to obtaining the optimal solution using off-the-shelf solvers. Second, we modeled the problem as a multi-agent Markov decision process (MMDP), in order to solve it efficiently using deep reinforcement learning (DRL) techniques, specifically, the deep Q-networks (DQN). The proposed DRL-based placement and migration scheme ensures service continuity under high mobility constraints and offer a reduced total service latency as well as the additional operational costs associated with the migration. The proposed scheme performs proactive placement of the requested services while considering the mobility of vehicles, the required amounts of computational and communication resources, and the overall migration costs.

To summarize, the main contributions of this paper are synthesized as follows:
\begin{itemize}
    \item We formulate the service placement and migration problem as a non-linear program to minimize the total service latency (including the computing latency and the communication latency) and the cross-edge operational costs. 
    \item We propose an MMDP framework that helps solving the problem in a distributed and scalable manner.
    %\item We leverage proposed scheme allows the SDN controller in charge of performing MEC nodes' resource provisioning for different services by solving an associated multi-agent MDP model dependently on the state of each MEC node. And we propose a multi-agent deep Q-learning (DQL) approach, specifically the deep Q-network to allocate the computational resources to services being migrated under low latency constraints.
    \item We leverage DRL techniques to provide efficient solution to the MMDP model. Specifically, we propose a deep Q learning (DQL)-based solution that uses double Q network and replay buffer to improve the learning outcome. 
    \item We evaluate the performance of the proposed DRL-based scheme and compare it to the optimal solution obtained by the CPLEX solver and we show that the proposed solution achieves close-to-optimal performance.
\end{itemize}

The remainder of this paper is structured as follows. In Section II, we present the system model and the problem formulation of the service placement and migration problem. In Section III, we present the proposed multi-agent DQL-based solution. The performance of the proposed solution is evaluated in Section IV. Last but not least, the related works are discussed in Section V. Finally, the paper is concluded in Section VI.

\section{System Model and Problem Formulation}

\subsection{System Model}
\begin{figure}[t]
    \centering
    \includegraphics[width=.7\linewidth]{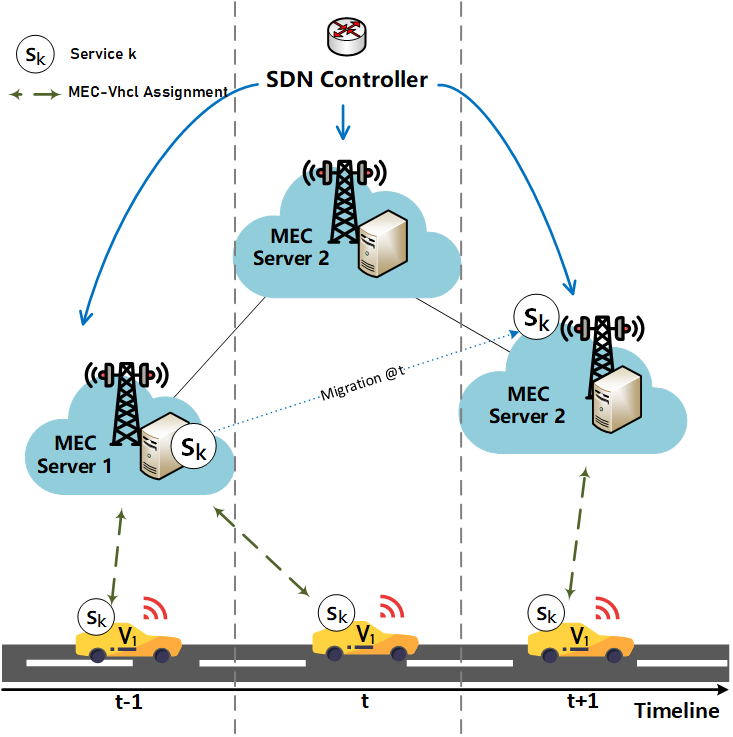}
    \caption{Illustration of the system model}
    \label{fig:arch}
\end{figure}

We consider an SDN-enabled MEC architecture covered with a set of gNodeBs (gNBs), each is equipped with a MEC server $n\in\mathcal{N}\coloneqq\{1, 2, \dots, N\}$ that is connected to one gNB via high speed local-area network as illustrated in Fig. \ref{fig:arch}. There are $K$ mobile users (or interchangeably called vehicles) demanding services from the MEC servers and are denoted by the set $\mathcal{K}\coloneqq\{1, 2, \dots, K\}$. Each vehicle $k$ requests some service to fulfill its requirements. Without loss of generality, we assume that all vehicles request the same vehicular service\footnote{The case of multiple services will be considered in our future work where network slicing will be integrated into our system model.} (e.g., an infotainment-related service). Similar to previous works~\cite{9220170, 9014146, URGAONKAR2015205, 8647545}, we consider a MEC-based device-oriented service model contrary to the traditional cloud-based application-oriented service model. In other words, a dedicated container or virtual machine is assigned the vehicular service as well as the applications' environment, which are executed on each vehicle rather than on each application. An SDN controller is assumed to be placed on the cloud layer where it acts as a central controller for information exchange between vehicles. Time is discrete and is divided into a set of $T$ time-slots denoted by the set $\mathcal{T}=\{1, 2, \dots, T\}$. At each time-slot $t\in\mathcal{T}$, each vehicle $k\in\mathcal{K}$ requests the vehicular service from the MEC node $n\in\mathcal{N}$. 

% We consider a single agent MDP where the SDN controller plays the role of an intelligent agent, called hereinafter the SDN agent. The actions available for the SDN agent are given by the set $\mathcal{A}$, where an action from $\mathcal{A}$ is given by the placement of the service of user $u$ at some MEC node $m$ at time-slot $t$.

The objective of this work is to guarantee the minimum quality of service (QoS) requirements of the vehicles while considering their erratic mobility and the computing and communication resources of MEC servers. To do so, the requested vehicular service should be placed and migrated across different MEC servers depending on the vehicles mobility patterns. In this work, we consider an hybrid centralized-distributed architecture where (i) each MEC server plays the role of an agent that makes its service placement and migration decisions independently of other MEC servers, and (ii) once each MEC agent makes its decision, it communicates it to the SDN controller that plays the role of a central agent to coordinate the decisions of all MEC servers. 

The considered QoS is represented by the vehicular service latency that includes (i) the communication delay that is incurred by the transmission between a vehicle and a MEC server, and (ii) the computing delay that depends on the processing capability of the MEC server as well as the size of the vehicle's request. %and (iii) the migration cost that depends on the bandwidth and the energy costs~\cite{8463562}.
\subsubsection{Communication Delay}
When a vehicle $k$ requests from a MEC server $n$ the vehicular service, the transmission between $k$ and $n$ depends mainly on the wireless environment and on the size of the requested service. The channel power gain between vehicle $k$ and MEC server $n$ at time-slot $t$ is denoted by $g_{kn}^t$, which includes the small-scale fading as well as the large-scale fading. To simplify the analysis, we assume that the total available bandwidth is divided equally between the MEC servers and each MEC server allocates its bandwidth to the vehicles in an orthogonal manner. Accordingly, the received signal to noise ratio between MEC server $n$ and vehicle $k$ at time-slot $t$ is given by 
\begin{align}
    \label{snr}
    \gamma_{kn}^t\coloneqq\dfrac{p_ng_{kn}^t}{\sigma^2},
\end{align}
where $p_n$ is the transmit power of MEC server $n$ and $\sigma^2$ is the power of the noise. The achieved data rate can be given as follows
\begin{align}
    \label{rate}
    \Gamma_{kn}^t\coloneqq w_n\lg\left(1+\gamma_{kn}^t\right),\qquad\text{ [in bits/sec]}
\end{align}
where $w_n$ is the allocated bandwidth of MEC server $n$. Consequently, the communication delay between MEC server $n$ and vehicle $k$ at time-slot $t$ is given by:
\begin{align}
    \label{commund}
    d_{kn}^t\coloneqq\dfrac{s_k}{\Gamma_{kn}^t},\qquad\text{ [in sec]}
\end{align}
where $s_k$  is the size of the requested service of vehicle $k$.

\subsubsection{Computing Delay}
The computing delay depends on the processing capacity of each MEC server $n$, on the total vehicles sharing MEC server $n$, and on the requested computing capacity of the vehicular service of vehicle $k$ at time-slot $t$. More precisely, the computing delay between MEC server $n$ and vehicle $k$ at time-slot $t$ is given as follows~\cite{8463562}:
\begin{align}
    \label{compd}
    c_{kn}^t\coloneqq c_k^tN_n^t/F_n,\qquad\text{ [in sec]}
\end{align}
where $c_k^t$ denotes the amount of computing capacity [in CPU cycles] required by the requested vehicular service of vehicle $k$ at time-slot $t$. The computing capacity of MEC server $n$ [in CPU cycles/sec] is given by $F_n$ and the number of vehicles placed on MEC server $n$ at time-slot $t$ is given by $N_n^t$.

Besides the QoS requirements, placing and migrating the vehicular service across multiple MEC servers incur additional operational costs related, for example, to the energy consumption and the bandwidth usage. For this reason, we consider the migration cost as an important factor into the design of our service migration solution. 
\subsubsection{Migration Cost}
Due to the cross-edge migration, additional operational costs are incurred by the service migration. These costs include energy consumption, expensive wide-area network bandwidth usage, etc.~\cite{8463562}. To make the operational cost model general, we use $m_{n'n}^{kt}$ to denote the cost of migrating the vehicular service of vehicle $k$ from MEC server $n'$ to MEC node $n$ at time-slot $t$. Obviously, we assume that $m_{n'n}^{kt} = 0$, for all $n' = n$ and for all $k,t$.

\subsection{Problem Formulation}
To guarantee the required QoS (communication and computing delays) and the migration cost, the optimization problem is formulated as a multi-objective optimization problem where the aim is to optimize the communication delay, the computing delay as well as the migration costs. To simplify the resolution of this multi-objective problem, we transform the multi-objective problem into a single objective one by introducing the weights $\lambda_i$ for $i\in\{1,2,3\}$. The formulated problem is written as a nonlinear integer program (NLP) as follows. 

\begin{problem}\label{pb:1}
  \begin{alignat}{2} &\Minimize_{\mathbf{x}}&\qquad&\lambda_1C(\mathbf{x})+\lambda_2D(\mathbf{x})+\lambda_3M(\mathbf{x})\label{obj:1}\\
  & \subjto
  & & x_{kn}^t\in\{0,1\}, \quad \forall k,n,t, \label{C:def1}\\
  & & & \sum_{n=1}^{N}x_{kn}^t=1, \quad \forall k,t, \label{C:oneMEC}
 \end{alignat}
\end{problem}
where the variables $x_{kn}^t=1$ if and only if the vehicular service requested by vehicle $k$ is placed at MEC server $n$ at time-slot $t$. We denote by $\mathbf{x}$ the multidimensional notation of the variables $x_{kn}^t$, i.e., $\mathbf{x}=[x_{kn}^t]$. The objective function in~\eqref{obj:1} is a linear combination of the communication delay, the computing delay, and the migration cost. Constraints~\eqref{C:def1} guarantee that the variables $x_{kn}^t$ are binary. Constraints~\eqref{C:oneMEC} guarantee that the vehicular service requested by vehicle $k$ at time-slot $t$ is placed at one and only one MEC server.

The total computing delay $C(\mathbf{x})$ is defined as follows:
\begin{align}\label{comput_delay}
    C(\mathbf{x})\coloneqq\sum_{n=1}^N\sum_{k=1}^K\sum_{t=1}^Tx_{kn}^tN_n^tc_k^t/F_n,
\end{align}
where $c_k^t$ denotes the required amount of computing capacity [in CPU cycles] of the vehicular service requested by vehicle $k$ at time-slot $t$ and $F_n$ denotes the maximum computing capacity of MEC server $n$ [in CPU cycles/sec]. The term $N_n^t$ denotes the number of services placed at MEC server $n$, i.e.,
\begin{align}\label{num_services}
    N_n^t\coloneqq\sum_{k=1}^Kx_{kn}^t.
\end{align}

The total communication delay $D(\mathbf{x})$ is defined as follows:
\begin{align}\label{commun_delay}
    D(\mathbf{x})\coloneqq\sum_{n=1}^N\sum_{k=1}^K\sum_{t=1}^Tx_{kn}^td_{kn}^t,
\end{align}
where $d_{kn}^t$ denotes the computing delay between MEC server $n$ and the vehicle $k$ at time-slot $t$ (see~\eqref{commund}). %the general communication delay between MEC node $i$ and a user when its corresponding service $k$ is placed on MEC node $i$ at time-slot $t$. It mainly depends on the link bandwidth between the MEC node and the user and the amount of data transferred to the user. We use a general temr $d_i^k(t)$ to make our problem generic and independent to a specific transmission model.

Finally, the total migration cost $M(\mathbf{x})$ is defined as follows:
\begin{align}\label{mig_cost}
    M(\mathbf{x})\coloneqq\sum_{n=1}^N\sum_{n'=1}^N\sum_{k=1}^K\sum_{t=1}^Tx_{kn'}^{t-1}x_{kn}^tm_{n'n}^{kt},
\end{align}
where the migration cost $m_{n'n}^{kt}$ is used to denote the cost of migrating the service $k$ from MEC server $n'$ to MEC server $n$ at time-slot $t$. It is clear that the cost is counted inside the summation only if both $x_{kn'}^{t-1}$ and $x_{kn}^t$ are equal to one, i.e., $x_{kn'}^{t-1}=x_{kn}^t=1$, which means that the requested service of vehicle $k$ is placed at MEC server $n'$ at time-slot $t-1$ and is placed at MEC server $n$ at time-slot $t$. This costs includes bandwidth costs incurred by cross-edge migration (e.g., wide-area network bandwidth usage costs) as well as energy costs caused by increased energy consumption of network devices such as routers. To make the model general, we use a general cost term $m_{n'n}^{kt}$ as done in~\cite{8463562}.

In order to make the problem more tractable, we linearize the objective function given in~\eqref{obj:1}. The non-linearity of~\eqref{pb:1} comes from the functions $C(\mathbf{x})$ and $M(\mathbf{x})$ due to the multiplication of binary variables. To linearize $M(\mathbf{x})$, we introduce a new binary variable called $z_{n'n}^{kt}=x_{kn'}^{t-1}x_{kn}^t$. It is clear that this new z-variable is positive if and only if each term of the product of the x-variables is positive. In other words, $z_{n'n}^{kt}=1\iff x_{kn'}^{t-1}=x_{kn}^t=1$. This means that we must add the following two constraints to force the z-variable to be zero whenever $x_{kn'}^{t-1}$ or $x_{kn}^t$ is zero:
\begin{align}\label{cns1_zx1}
    z_{n'n}^{kt}\leq x_{kn'}^{t-1},\forall k,n,n',t>1,
\end{align}
and
\begin{align}\label{cns1_zx2}
    z_{n'n}^{kt}\leq x_{kn}^t,\forall k,n,n',t.
\end{align}

It remains to enforce the constraints that if both $x_{kn'}^{t-1}$ and $x_{kn}^t$ are equal to one, then the z-variable is one. This can be written as follows:
\begin{align}\label{cns1_zx3}
    z_{n'n}^{kt}\geq x_{kn'}^{t-1}+x_{kn}^t-1,\forall k,n,n',t>1.
\end{align}
Thus, the total migration cost can be rewritten as follows:
\begin{align}\label{mig_cost2}
    M(\mathbf{z})=\sum_{n=1}^N\sum_{n'=1}^N\sum_{k=1}^K\sum_{t=1}^Tz_{n'n}^{kt}m_{n'n}^{kt},
\end{align}
where $\mathbf{z}$ denotes the multidimensional notation of the variables $z_{n'n}^{kt}$, i.e., $\mathbf{z}=[z_{n'n}^{kt}]$.

Now, to linearize $C(\mathbf{x})$, we let $y_k^t$ denote the quantity $\sum_{n=1}^Nx_{kn}^tN_n^tc_k^t/F_n$, i.e.,
\begin{align}\label{y1}
    y_k^t\coloneqq \sum_{n=1}^Nx_{kn}^tN_n^tc_k^t/F_n,\forall k,t.
\end{align}
Thus, the total computing delay $C(\mathbf{y})$ can be rewritten as follows:
\begin{align}\label{comput_delay2}
    C(\mathbf{y})=\sum_{k=1}^K\sum_{t=1}^Ty_k^t,
\end{align}
where $\mathbf{y}$ denotes the multidimensional notation of the variables $y_{k}^t$, i.e., $\mathbf{y}=[y_{k}^t]$. Now, we have to enforce that the following constraints:
\begin{align}
    x_{kn}^t=1 \Rightarrow  y_k^t=N_n^tc_k^t/F_n.
\end{align}
These are indicator constraints that can be easily implemented in the off-the-shelf solvers such as CPLEX or Gurobi. Nonetheless, they can be easily transformed to linear constraints using the big-M method~\cite{7460142, }.

\section{Proposed Solution}
In this section, we propose a deep reinforcement learning (DRL) based approach to obtain an efficient solution to the service placement and migration problem defined in \eqref{pb:1}. The proposed approach places the vehicular service requested by the vehicles in the appropriate MEC servers to ensure the continuity of services under the mobility constraint of vehicles while reducing the communication latency, the computing latency as well as the migration costs of the requested service. 

We use deep Q-learning (DQL)~\cite{mnih2015human}---one of the most popular DRL algorithm---to efficiently solve the service placement problem in the MEC-enabled vehicular network. DQL combines Q-learning with deep neural network (DNN). It takes as input the observed state of the environment and returns as output the Q-value of all possible actions. DQL has two main phases, namely the training phase and the inference phase. In the training phase, the agent trains a DNN, called deep Q-network (DQN), in an offline manner. In the inference phase, the agent takes actions in an online manner based on the trained DQN. Before describing each phase of the proposed DQL algorithm, we model, first, the problem as a Markov decision process (MDP).

\subsection{The MDP Formulation}
We consider a multi-agent MDP where each MEC server $n$ acts as an independent agent, called herein after MEC agent $n$. At time-slot $t$, each MEC agent $n$ can decide either to place and instantiate the vehicular service requested by vehicle $k$ or not. The key elements of the multi-agent MDP are defined as follows: 

\subsubsection{The State Space}
At time-slot $t$, the observed state by the MEC agent $n$, denoted by $\mathcal{S}_n^t$, mainly depends on the current vehicular environment. It includes the current positions of the vehicles, their velocities, their directions, and their service requirements (including the wireless channel gains and the distances between MEC agent $n$ and the vehicles). Note that there are as many states as there are time-slots, i.e., every time-slot corresponds to a state. In addition, a transition from one state to the next happens according to the mobility model of the vehicles.

\subsubsection{The Action Space}
The action set of each MEC agent $n$ at time-slot $t$ is given by the set $\mathcal{A}_n^t\coloneqq\{0,1\}^K$. Indeed, an action $\boldsymbol{a}_n^t\in\mathcal{A}_n^t$ is given by the row vector $[a_{1n}^t,a_{2n}^t,\dots,a_{Kn}^t]$, where each element $a_{kn}^t$ corresponds to the decision to place the service $k \in \mathcal{K}$ at MEC server $n$, all happening at time-slot $t$. Note that the variable $a_{kn}^t$ and $x_{kn}^t$ defined in~\eqref{pb:1} means essentially the same thing but to remove any possible confusion between the optimization variable $x_{kn}^t$ and the MDP action $a_{kn}^t$ we use two different notations. Each MEC agent $n$ communicates its chosen action to the SDN controller to form a global action $\boldsymbol{a}^t\coloneqq[\boldsymbol{a}_1^t,\boldsymbol{a}_2^t,\ldots,\boldsymbol{a}_N^t]$. Then, the SDN controller verifies if the individual actions of the MEC agents are feasible or not according to the constraints of~\eqref{pb:1}, i.e., the individual action $\boldsymbol{a}_n^t$ of MEC agent $n$ is considered feasible if it meets the constraints of~\eqref{pb:1}.

\subsubsection{The Reward Function}
A MEC agent $n$ chooses an action $\boldsymbol{a}_n^t\in\mathcal{A}_n^t$ at time-slot $t$ and receives a reward $R_n^t$. Since we seek to minimize the overall vehicular service latency requested by the vehicles, the objective of MEC agent $n$ must be related to the sum-latency of the services it hosts. In other words, we define the reward $R_n^t$ of MEC agent $n$ at time-slot $t$ in relation with how the placement of requested service at $n$ affects the latency of the system. Therefore, the SDN controller calculates the individual reward of MEC agent $n$ as follows:

\begin{align}\label{MECReward}
    R_n^t\coloneqq 
    \begin{cases}
    \lambda_1C_n^t + \lambda_2D_n^t + \lambda_3M_n^t, & \text{if } \boldsymbol{a}_n^t \text{ is feasible}\\
    -1, & \text{if } \boldsymbol{a}_n^t \text{ is not feasible},
    \end{cases}
\end{align}
where $C_n^t = \sum_{k=1}^K a_{kn}^tN_n^tc_k^t/F_n$ is the computation delay, $D_n^t = \sum_{k=1}^Ka_{kn}^td_{kn}^t$ is the communication delay, and $M_n^t = \sum_{\substack{n'=1\\n'\neq n}}^N\sum_{k=1}^Ka_{kn'}^{t-1}a_{kn}^tm_{n'n}^{kt}$ is the migration cost of MEC agent $n$ at time-slot $t$. If the action chosen by MEC agent $n$ at time-slot $t$ is not feasible, this MEC agent should be penalized with a negative reward $R_n^t = -1$ to prompt it to not choose this action in future steps. 

\subsection{The Training Phase of DQL}
\begin{algorithm}[!t]
\SetAlgoLined
\SetKwInOut{Initialization}{Initialization}
\KwIn {Agents and environment}
\textbf{Output:} Trained DDQNs

\Initialization {Generate vehicles and network parameters;
                \\ Initialize the DDQN of each agent $n$;}
\For{Episode $e$}{
	Reset and build the agents' environment\;
	\For{Time-slot $t$}{
		\For{each MEC $n$}{
			Observe the environment \;
			Choose an action $a_n^t$ using $\epsilon$-greedy\;
		}
		The SDN controller obtains the global action \;
		The SDN controller calculates the individual reward of each agent\; 
		\For{each MEC $n$}{
			Receive the individual reward from the SDN controller\;
			Observe the next state of the environment \;
			Store the experience $Exp_n^t$ in the replay buffer $\mathcal{M}_n$\;
			\If{batch size} {
			    Sample a mini-batch from $\mathcal{M}_n$\;
				Do a mini-batch training\;}
            \If{target step}{
                Update the target network parameters $\theta^{-}_n$\;}
		}
	}
}
    \caption{The Training Phase of DQL}
    \label{alg:train}
\end{algorithm}
In general, DQN approximates the Q-values $Q(s,a,\theta)$ of each state-action pair $(s, a)$ using a DNN, where $\theta$ represents the parameters of the Q-network. Since we propose a multi-agent MDP, the proposed DQL algorithm will be a multi-agent algorithm in which each MEC agent will have its own DQN to be approximated and trained. When there is no confusion, we omit the index $n$ from the DQN of MEC agent $n$. In addition, the training process of the DNN uses the experience replay memory mechanism. This mechanism helps in creating a dataset to train the DNN once in a while by storing each MEC agent experience into a replay buffer. This experience essentially includes the current state, the next transition state, the chosen action and the received reward. Then, each MEC agent randomly chooses a set of samples from its replay buffer to perform the learning process. The experience replay memory mechanism not only allows the MEC agent to learn from the past experiences, but also to provide uncorrelated data as inputs which breaks undesirable temporal correlations. However, DQN is known to overestimate the Q-values of stat-action pairs under certain conditions, which harms the performances. To overcome this issue, double DQN (DDQN)~\cite{van2016deep} is proposed which reduces the overestimation and makes the training process faster and more reliable. Indeed, DDQN uses two DNNs, called the main Q-network and the target Q-network. The former is used to compute the Q-values $Q(s,a,\theta)$ while the latter is used to provide the target Q-values $Q(s,a,\theta^{-})$ to train the parameters $\theta$ of the main Q-network. The training phase of our proposed multi-agent DQL algorithm is presented in Algorithm~\ref{alg:train}, where each MEC agent $n\in\mathcal{N}$ trains its own DDQN.

The training phase of the DQL algorithm requires as input the vehicular environment which includes the vehicles, the requested services, the MEC servers, the computing capacity of MEC servers. It returns the trained DDQN of each MEC agent as output. The DDQNs are trained simultaneously. The training begins by generating the vehicles parameters and the network parameters. The vehicles parameters include the position, the velocity and the requested service of each vehicle. The network parameters include the computing capacity of each MEC server. Then, the DQL algorithm initializes the DDQN parameters of each MEC agent. Next, it iterates the episodes. For each episode, the environment of each MEC agent is built by updating the position of the vehicles according to the mobility model and generating other network parameters. For each episode, the training continues for a period of time-slots (or steps). In each step $t$, each MEC agent $n$ observes the current state of its environment and chooses an action from its action space $\mathcal{A}_n^t$. To select an action, the MEC agent uses the $\epsilon$-greedy policy. With this policy, an action is chosen randomly with probability $\epsilon$. Once all MEC agents select their actions, each of them communicates its action to the SDN controller to construct the global action $\boldsymbol{a}^t$. The SDN controller uses the constructed global action to verify its feasibility and calculate the individual reward of each MEC agent. Then, each MEC agent $n$ receives its individual reward $R_n^t$ from the SDN controller and moves to the next state. The obtained experience, denoted by $Exp_n$, is stored by the MEC agent $n$ in its replay buffer $\mathcal{M}_n$. When the replay buffer contains enough experiences, i.e., a certain batch size is respected, each MEC agent randomly samples a mini-batch to create a training dataset. The latter is used by the MEC agent to perform the training process. In the training process, each MEC agent seeks to minimize a loss function, given by:
\begin{align}
    \label{lossF}
    L_n^t(\theta_n)=\mathbbm{E}[(y_n - Q(\mathcal{S}_n^t,\boldsymbol{a}_n^t;\theta_n))^2],
\end{align}
where $Q(\mathcal{S}_n^t,\boldsymbol{a}_n^t;\theta_n)$ is the Q-value of action $\boldsymbol{a}_n^t$ given the state $\mathcal{S}_n^t$ which is calculated using the main Q-network with parameters $\theta_n$; $y_n$ is the target Q-value, which calculated using the target Q-network with parameters $\theta^{-}_n$ and it is given as follows:
\begin{align}
    \label{targetV}
    y_n=R_n^t + \gamma Q(\mathcal{S}_n^t,\underset{\boldsymbol{a}_n^t}{max} \,\{ Q(\mathcal{S}_n^t,\boldsymbol{a}_n^t;\theta_n)\};\theta^{-}_n),
\end{align}
where $0 \leq \gamma \leq 1 $ is the discount factor.

To update the parameters $\theta_n$ of the main Q-network, MEC agent $n$ performs a gradient descent step. Finally, each MEC agent updates the parameters $\theta^{-}_n$ of its target Q-network at a fixed target step by copying the parameters of the main Q-network.

\subsection{The Inference Phase of DQL}
The inference phase of DQL is presented in Algorithm~\ref{alg:implem}. Once the trained DDQNs are obtained, each MEC agent uses its optimal DDQN parameters to find an appropriate placement of the requested service by the vehicles. In detail, at the beginning of each episode the environment of each MEC agent is built. Then, for each step $t$, each MEC agent observes the current state of its environment and selects an action that maximizes its Q-value according to its trained DDQN. Based on the selected actions of all MEC agents, the SDN controller finds the overall communication delay, computing delay and migration costs and thus we obtain a solution to problem~\eqref{pb:1}.

\begin{algorithm}[t]
\SetAlgoLined
\SetKwInOut{Initialization}{Initialization}
\KwIn {The trained DDQNs}
\textbf{Output:} Placement of the vehicular service of each vehicle

\Initialization {Load the DDQN of each agent $n$;}
\For{Episode $e$}{
	Reset and build the environment\;
	\For{Step $t$}{
		\For{each MEC $n$}{
			Observe the environment \;
			Choose an action $a_n^t$ that maximize the Q-value of the tained DDQN of $n$\;
		}
		The SDN controller obtains the global action\;
	}
	The SDN controller calculates the objective function as in~\eqref{obj:1}\;
}
    \caption{The Inference Phase of DQL}
    \label{alg:implem}
\end{algorithm}

\section{Simulation Results}
We consider a MEC-enabled vehicular network where three gNBs that are attached to three MEC servers are deployed over a highway as shown in Fig. 1. The gNBs are located randomly along the highway. We assume that the three MEC servers are deployed along the highway in a triangular fashion as depicted in Fig. 1, where the distance between MEC 1 and MEC 2 and the distance between MEC 2 and MEC 3 is equal to 2000 m and the distance between MEC 1 and MEC 3 is equal to 4000 m. The vehicles are drawn randomly in the highway that is modelled as a rectangle of length 5000 m and width 18 m with two forward lanes and two backward lanes. The vehicles move with a randomly-chosen fixed speed from the range of $[60, 110]$ km/h and once a vehicle reaches the boundary of the highway, it reappears in the opposite side. For simplicity, all vehicles keep moving with constant speeds with acceleration, i.e., once the random speed of a vehicle is chosen, the latter keeps moving with that speed during the entire simulation period.  

The proposed multi-agent DQL algorithm is trained on an computer with an Intel Core i7-10750H CPU, 16GB RAM and an nVidia GeForce GTX 2070 Super graphic card. The implementation is performed using Python and PyTorch. After performing hyper-parameters tuning, the following optimized parameters are set. Each DDQN consists of fully connected hidden neural network with two hidden layers of 256 neurons each. The discount factor is $\gamma=0.99$. The other DDQN and vehicular network para metes are presented in Table 1. To avoid the overestimation problem of the Q-value, the parameters of each DDQN network are copied into the parameters of the corresponding target DDQN every 1000 steps. According to the state of the art of deep learning models, Rectified Linear Unit (ReLU) function accelerates the learning process since it is not vanishing gradient.
\begin{table}[t]
\vspace{.2in}
    \renewcommand{\arraystretch}{1.3}
    \caption{Simulation parameters}
    \label{tab:sim-setup}
    \centering
    \begin{tabular}{|| l | l ||}
    \hline
        \textbf{Parameter} & \textbf{Value} \\ \hline\hline 
        Number of MEC servers &  $3$\\\hline
        Transmit power of each gNB & $30$ dBm\\\hline
        Migration cost & $\textsc{uniform}(0.2, 0.3)$\\\hline
        Number of vehicles &  $4$\\\hline
        Request size &  $\textsc{uniform}(50, 300)$ Kbits \\\hline
        Noise variance & $-174$ dBm/Hz\\ \hline
        Bandwidth & 10 MHz\\\hline
        Learning rate &  $3e-4$\\\hline
        Number of episodes & $3000$ \\\hline
        Discount factor &  $0.99$\\\hline
        Replay memory size &  $100000$\\\hline
        Mini-batch size & $1024$ \\\hline
        Target update interval &  $1000$\\\hline
        Loss function &  Minimum square error\\\hline
        Optimizer & Adam \\\hline
        Activation function & ReLU \\ \hline
    \end{tabular}
\end{table}

\begin{figure}[t]
  \centering
% \vspace{.2in}
  \includegraphics[width=.7\linewidth]{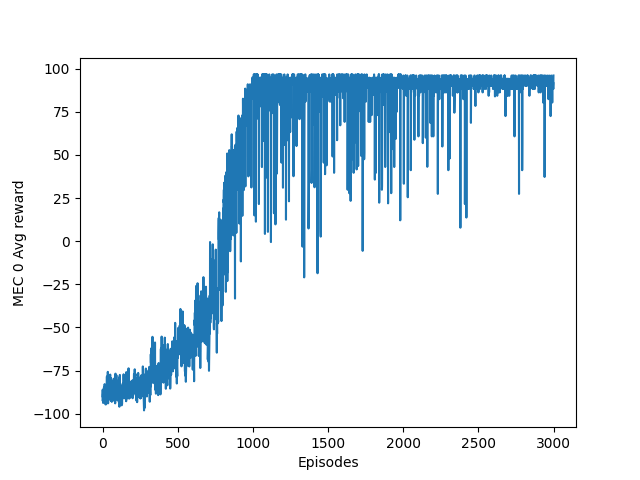}
  \caption{The training rewards for MEC server.}
  \label{fig:rewards}
\end{figure}

Fig.~\ref{fig:rewards} illustrates the average reward per episode of one MEC agent. It is clear that the reward improves with the training episodes as it increases when the number of episodes increases. This shows the effectiveness of the proposed DQL algorithm. We notice that the DQL algorithm converges at approximately 1000 episodes. In other words, the corresponding MEC agent converges to a good learning outcome, which implies that it will explore better actions. We can notice though that the reward converges while incurring large fluctuations which is mainly due to the high mobility scenario of the vehicular network.

\begin{figure}[b]
    \centering
    \includegraphics[width=.7\linewidth]{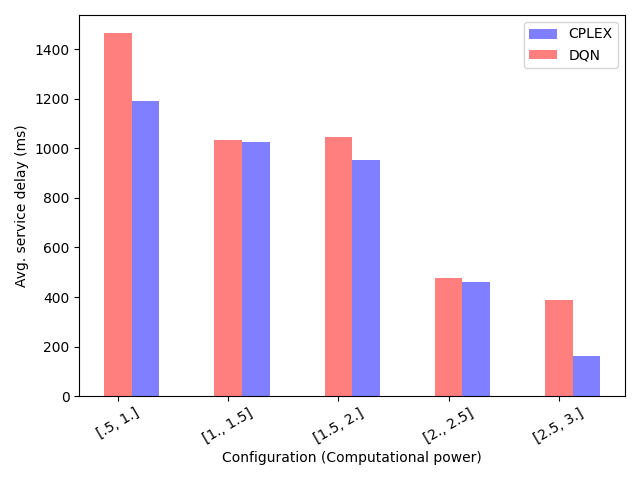}
    \caption{The objective function vs. the computational power of the MEC servers.}
    \label{fig:comp}
\end{figure}

\begin{figure}[b]
    \centering
    \includegraphics[width=.7\linewidth]{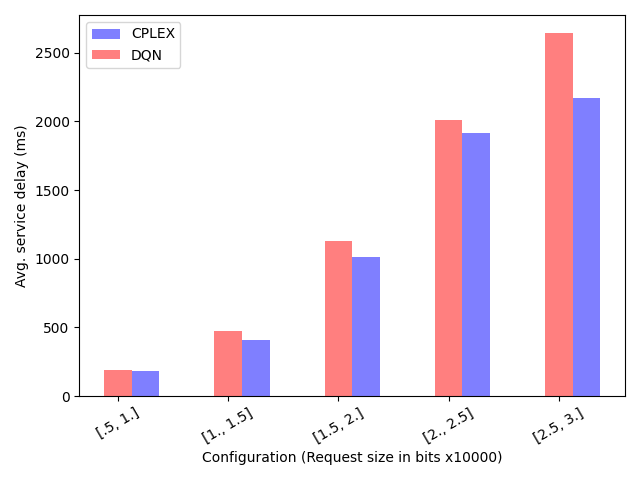}
    \caption{The objective function vs. the request sizes of the vehicles.}
    \label{fig:task}
\end{figure}

Fig. \ref{fig:comp} and Fig. \ref{fig:task} illustrate the average cost represented by the objective function~\eqref{obj:1} which measures the total service latency (the computing and communication latency) as well as the migration costs under two different configurations. The first configuration is the computational power configuration and it consists of varying the number of cores for the three MEC servers. We considered three MEC servers with $4$ cores each, or $8$ cores each, or $16$ cores each, or $32$ cores each, or $64$ cores each, with a fixed clock frequency of $2.5$ GHz. The second configuration is the request size configuration and it consists of varying the vehicles' request sizes, which we generate uniformly at random within a fixed interval as follows \textsc{uniform}(50, 100), $\textsc{uniform}(100, 150)$, $\textsc{uniform}(150, 200)$, $\textsc{uniform}(200, 250)$, $\textsc{uniform}(250, 300)$ Kbits.

Fig. \ref{fig:comp} shows the objective function while considering the computational power configuration. We can notice that with increasing the computational power, the average service latency as well as the migration costs are decreasing. Regardless of different number of cores, the proposed DQL approach performs close-to-the-optimal performance.
Fig. \ref{fig:task} presents the objective function while considering the request size configuration. The proposed approach is shown to approximate well the optimal solution in different scenarios of the request size configuration. The smaller the request size is, the smaller the total delay and the migration costs are. This is because (i) the DQL approach learns efficiently the appropriate placement of each vehicle's service, which helps in reducing the service latency and (ii) a small number of bits can be fulfilled easily by one MEC server without requiring to migrate to another MEC server, which helps in reducing the total migration costs. We notice that as the request sizes increase, the average service delay and the migration costs increase as well.

\section{Related Works}

The authors in \cite{haung2021qoe} propose a quality-of-experience (QoE)-aware scheme to ensure service continuity for mobile cloud computing environment. The scheme relays on the buffer-occupancy threshold policy that classifies the new arriving request from the mobile users. The proposed scheme protects the migrated service from traffic fluctuation. In addition, the cloud server can change the buffer threshold dynamically for different categories of requests. In \cite{8463562}, the authors proposed Follow-Me Chain algorithm to solve the problem of service function chaining (SFC). In particular, the work studied the problem of inter-MEC handoffs to offer a higher satisfaction for users in mobility scenarios. Such problem is NP-hard, and authors proposed an integer programming formulation that is solved by the Follow-Me Chain algorithm. The work in \cite{roy2020user} investigated the relocation problem of virtual network functions (VNF) within a cloud infrastructure under mobility and resource heterogeneity constraints. The authors studied in particular the impact of the relocation operation on the service delay and the number of VNF relocations (i.e., the number of time that a single VNF is being moved from a cloud to another). The problem of relocation was formulated as a mized integer linear programming problem and solved through a meta-heuristic approach, namely, the ant colony optimization technique. Within the same context, the authors in \cite{addad2018towards}  proposed an evaluation of three container-based schemes for VNF migration as a mechanism to guarantee the service continuity. In particular, the schemes consider two cases of mobility patterns, respectively, known a priori and unknown mobility patterns. For the known a priori pattern, temporary file system and disk-less-based migration are discussed, but the main focus was on the unknown mobility pattern, where authors proposed a solution that consists in storing the container's file system within the system images in a shared pool. The work in \cite{khan2021network} considered two main logical slices created over the same infrastructure, namely, an autonomous driving slice for safety messages, and an infotainment slice. The authors proposed a clustering method to partition vehicles to allocate slice leaders on each cluster. A slice leader is a serving entity using vehicle-to-vehicle (V2V) links to forward safety messages, subsequently the road side units (RSU) forward the infotainment service using the vehicle-to-infrastructure (V2I) links. In \cite{8676226}, the authors proposed an offline RL-based RAN slicing solution and a low-complexity heuristic algorithm, to satisfy communication resources requirements of different slices with the aim to maximize the resource utilization. The proposed approach ensures the resource availability to meet the different requirements of the slice's traffic. The authors assume that V2V communications are either in cellular (through gNBs) or in sidelink mode (through PC5 communication). In addition, in the sidelink mode each vehicle can multicast to multiple vehicles within the same cluster. Finally, the proposed RL approach is executed separately for each communication mode (i.e., uplink and downlink), which means that the RL is being executed twice.

Most of the literature studied hereabove, focus on the resource provisioning at the MEC sides independently with no consideration to the services migration problem in vehicular network, where factors such such as the mobility patterns, the services migration costs, and the services requirements need to be considered for a more efficient service placement schemes. Further, previous works consider only heuristic or meta-heuristic methods that focus on solving the placement problem in order to minimize only the latency without studying the cost of migration. The works that considered the service latency as well as the migration costs leverage simple algorithmic solutions without considering advanced machine learning approaches such as the one proposed in this paper. In this paper, we fill these gaps and we propose a service migration scheme based on DRL techniques in a MEC-enabled vehicular network aiming to minimize the total service latency and migration cost.

\section{Conclusion}
In this paper, we proposed a DRL-based scheme to solve the problem of vehicular service placement and migration in a MEC-based vehicular network. First, we formulated the problem as a nonlinear integer optimization problem to minimize the total latency (i.e., communication and computational delays ) plus migration costs in terms of energy consumption and bandwidth usage. To solve the optimization problem we used standard solvers such as CPLEX, and we linearize it to transform it into a linear integer optimization problem. Then, we formulate the problem as a multi-agent Markov decision process and develop a DRL-based method by exploiting the DQL algorithm to obtain an efficient and non-complex solution. The DQL algorithm uses double DQN and replay memory strategies to increase the learning accuracy and solve the Q-value overestimation problem. Finally, we have demonstrated through extensive simulations that the proposed DQL algorithm achieves near-optimal performance compared to the CPLEX solution.

\section*{Acknowledgment}
The authors would like to thank the Natural Sciences and Engineering Research Council of Canada, for the financial support of this research.
\bibliography{references} 

\begin{thebibliography}{10}
\providecommand{\url}[1]{#1}
\csname url@samestyle\endcsname
\providecommand{\newblock}{\relax}
\providecommand{\bibinfo}[2]{#2}
\providecommand{\BIBentrySTDinterwordspacing}{\spaceskip=0pt\relax}
\providecommand{\BIBentryALTinterwordstretchfactor}{4}
\providecommand{\BIBentryALTinterwordspacing}{\spaceskip=\fontdimen2\font plus
\BIBentryALTinterwordstretchfactor\fontdimen3\font minus
  \fontdimen4\font\relax}
\providecommand{\BIBforeignlanguage}[2]{{%
\expandafter\ifx\csname l@#1\endcsname\relax
\typeout{** WARNING: IEEEtran.bst: No hyphenation pattern has been}%
\typeout{** loaded for the language `#1'. Using the pattern for}%
\typeout{** the default language instead.}%
\else
\language=\csname l@#1\endcsname
\fi
#2}}
\providecommand{\BIBdecl}{\relax}
\BIBdecl

\bibitem{7835337}
A.~Rachedi \emph{et~al.}, ``Ieee access special section editorial: The plethora
  of research in internet of things (iot),'' \emph{IEEE Access}, vol.~4, pp.
  9575--9579, 2016.

\bibitem{8647858}
L.~Yala \emph{et~al.}, ``{Latency and Availability Driven VNF Placement in a
  MEC-NFV Environment},'' in \emph{Proc. IEEE Global Commun. Conf. (GLOBECOM)},
  2018, pp. 1--7.

\bibitem{9497103}
A.~Alalewi \emph{et~al.}, ``On 5g-v2x use cases and enabling technologies: A
  comprehensive survey,'' \emph{IEEE Access}, vol.~9, pp. 107\,710--107\,737,
  2021.

\bibitem{triwinarko2021phy}
A.~Triwinarko \emph{et~al.}, ``Phy layer enhancements for next generation v2x
  communication,'' \emph{Vehicular Communications}, vol.~32, p. 100385, 2021.

\bibitem{7585028}
M.~Azizian \emph{et~al.}, ``An optimized flow allocation in vehicular cloud,''
  \emph{IEEE Access}, vol.~4, pp. 6766--6779, 2016.

\bibitem{7840359}
S.~Wang \emph{et~al.}, ``{An Investigation Into the Use of Virtual Reality
  Technology for Passenger Infotainment in a Vehicular Environment},'' in
  \emph{Proc. IEEE Int. Conf. Adv. Mater. Sci. Eng. (ICAMSE)}, 2016, pp.
  404--407.

\bibitem{9003407}
H.~Khan \emph{et~al.}, ``{Enhancing Video Streaming in Vehicular Networks via
  Resource Slicing},'' \emph{IEEE Trans. Veh. Technol.}, vol.~69, no.~4, pp.
  3513--3522, 2020.

\bibitem{8004158}
M.~Azizian \emph{et~al.}, ``Vehicle software updates distribution with sdn and
  cloud computing,'' \emph{IEEE Communications Magazine}, vol.~55, no.~8, pp.
  74--79, 2017.

\bibitem{9326402}
A.~Abouaomar \emph{et~al.}, ``Resource provisioning in edge computing for
  latency-sensitive applications,'' \emph{IEEE Internet of Things Journal},
  vol.~8, no.~14, pp. 11\,088--11\,099, 2021.

\bibitem{filali2020preemptive}
A.~Filali \emph{et~al.}, ``{Preemptive SDN Load Balancing with Machine Learning
  for Delay Sensitive Applications},'' \emph{IEEE Trans. Veh. Technol.},
  vol.~69, no.~12, pp. 15\,947--15\,963, 2020.

\bibitem{8450284}
P.~A. Frangoudis \emph{et~al.}, ``Service migration versus service replication
  in multi-access edge computing,'' in \emph{2018 14th International Wireless
  Communications Mobile Computing Conference (IWCMC)}, 2018, pp. 124--129.

\bibitem{9318243}
Z.~Mlika \emph{et~al.}, ``{Network Slicing with MEC and Deep Reinforcement
  Learning for the Internet of Vehicles},'' \emph{IEEE Network}, pp. 1--7,
  2021, {E}arly Access.

\bibitem{abouaomar2021service}
A.~Abouaomar \emph{et~al.}, ``{Service Function Chaining in MEC: A Mean-Field
  Game and Reinforcement Learning Approach},'' 2021.

\bibitem{abouaomar21globecom}
A.~{Abouaomar} \emph{et~al.}, ``{Mean-Field Game and Reinforcement Learning MEC
  Resource Provisioning for SFC},'' in \emph{IEEE GLOBECOM Conference}, 2021,
  pp. 1--6.

\bibitem{8290681}
A.~Aissioui \emph{et~al.}, ``On enabling 5g automotive systems using follow me
  edge-cloud concept,'' \emph{IEEE Transactions on Vehicular Technology},
  vol.~67, no.~6, pp. 5302--5316, 2018.

\bibitem{8737560}
T.~Ouyang \emph{et~al.}, ``{Adaptive User-Managed Service Placement for Mobile
  Edge Computing: An Online Learning Approach},'' in \emph{Proc. IEEE Conf.
  Comput. Commun. (INFOCOM)}, 2019, pp. 1468--1476.

\bibitem{8629587}
Z.~Ennya \emph{et~al.}, ``Computing tasks distribution in fog computing:
  Coalition game model,'' in \emph{2018 6th International Conference on
  Wireless Networks and Mobile Communications (WINCOM)}, 2018, pp. 1--4.

\bibitem{8463562}
T.~Ouyang \emph{et~al.}, ``{Follow Me at the Edge: Mobility-Aware Dynamic
  Service Placement for Mobile Edge Computing},'' \emph{IEEE J. Sel. Areas
  Commun.}, vol.~36, no.~10, pp. 2333--2345, 2018.

\bibitem{9220170}
A.~Tak \emph{et~al.}, ``Federated edge learning: Design issues and
  challenges,'' \emph{IEEE Network}, vol.~35, no.~2, pp. 252--258, 2021.

\bibitem{9014146}
A.~Abouaomar \emph{et~al.}, ``A resources representation for resource
  allocation in fog computing networks,'' in \emph{2019 IEEE Global
  Communications Conference (GLOBECOM)}, 2019, pp. 1--6.

\bibitem{URGAONKAR2015205}
R.~Urgaonkar \emph{et~al.}, ``{Dynamic Service Migration and Workload
  Scheduling in Edge-Clouds},'' \emph{Performance Evaluation}, vol.~91, pp.
  205--228, 2015.

\bibitem{8647545}
A.~{Abouaomar} \emph{et~al.}, ``{Matching-Game for User-Fog Assignment},'' in
  \emph{Proc. IEEE Global Commun. Conf. (GLOBECOM)}, 2018, pp. 1--6.

\bibitem{7460142}
Z.~Mlika \emph{et~al.}, ``{User–Base-Station Association in HetSNets:
  Complexity and Efficient Algorithms},'' \emph{IEEE Trans. Veh. Technol.},
  vol.~66, no.~2, pp. 1484--1495, 2017.

\bibitem{mnih2015human}
V.~Mnih \emph{et~al.}, ``{Human-Level Control Through Deep Reinforcement
  Learning},'' \emph{Nature}, vol. 518, no. 7540, pp. 529--533, 2015.

\bibitem{van2016deep}
H.~Van~Hasselt \emph{et~al.}, ``{Deep Reinforcement Learning with Double
  Q-Learning},'' in \emph{Proc. AAAI Conf. Artif. Intell.}, vol.~30, no.~1,
  2016.

\bibitem{haung2021qoe}
Y.-R. Haung, ``{A QoE-Aware Strategy for Supporting Service Continuity in an
  MCC Environment},'' \emph{Wireless Pers. Commun.}, vol. 116, no.~1, pp.
  629--654, 2021.

\bibitem{roy2020user}
P.~Roy \emph{et~al.}, ``{User Mobility and Quality-of-Experience Aware
  Placement of Virtual Network Functions in 5G},'' \emph{Comput. Commun.}, vol.
  150, pp. 367--377, 2020.

\bibitem{addad2018towards}
R.~A. Addad \emph{et~al.}, ``{Towards a Fast Service Migration in 5G},'' in
  \emph{Proc. IEEE Conf. Standards Commun. Netw. (CSCN)}, 2018, pp. 1--6.

\bibitem{khan2021network}
H.~Khan \emph{et~al.}, ``{Network Slicing for Vehicular Communication},''
  \emph{Trans. Emerg. Telecommun. Technol.}, vol.~32, no.~1, p. e3652, 2021.

\bibitem{8676226}
H.~D.~R. {Albonda} \emph{et~al.}, ``{An Efficient RAN Slicing Strategy for a
  Heterogeneous Network With eMBB and V2X Services},'' \emph{IEEE Access},
  vol.~7, pp. 44\,771--44\,782, 2019.

\end{thebibliography}
\bibliographystyle{IEEEtran}

\end{document}